\documentclass[twocolumn,aps,showpacs,floatfix,superscriptaddress]{revtex4}
\pdfoutput=1 %This is for the ArXiv submission only
\usepackage{amsmath,amssymb,eucal,graphicx,float,epstopdf}

\begin{document}
\title{Exclusion Processes with Avalanches}
\author{Uttam Bhat}
\affiliation{Department of Physics, Boston University, Boston, Massachusetts 02215, USA}
\author{P. L. Krapivsky}
\affiliation{Department of Physics, Boston University, Boston, Massachusetts 02215, USA}
\begin{abstract}
In an exclusion process with avalanches, when a particle hops to a neighboring empty site which is adjacent to an island the particle on the other end of the island immediately hops and if it joins another island this triggers another hop. There are no restrictions on the length of islands and the duration of the avalanche. This process is well-defined in the low-density region, $\rho<\frac{1}{2}$. We describe the nature of steady states (on a ring) and determine all correlation functions. For the asymmetric version of the process, we compute the steady state current, and we describe shock and rarefaction waves which arise in the evolution of the step-function initial profile. For the symmetric version, we determine the diffusion coefficient and examine the evolution of a tagged particle. 
\end{abstract}


\maketitle

\section{Introduction}

Lattice models which are endowed with conservative stochastic dynamics are known as lattice gases. Lattice gases were originally introduced, using the language of spin-exchange dynamics, by Kawasaki \cite{K66}, and they have played a crucial role in the following development of non-equilibrium statistical mechanics, see e.g. \cite{Spohn91,SZ95,D98,KL99,S00,BE07,D07,book} and references therein. One does not need to go to high dimensions to observe interesting behaviors in lattice gases---dynamics and non-equilibrium steady states are surprisingly rich already in one-dimensional lattice gases. 

The simple exclusion process (SEP) is perhaps the most well-known and widely studied interacting lattice gas. In the SEP, each site is either occupied by a particle or empty, and particles undergo nearest-neighbor hopping; only hops to empty sites are allowed and therefore particles interact only through the exclusion property. Two most popular versions, the symmetric simple exclusion process (SSEP) when hopping is symmetric and the totally asymmetric simple exclusion process (TASEP) when hopping is only in one direction, have been thoroughly investigated (see \cite{D98,KL99,S00,BE07,D07}). The simplest one-dimensional setting is a ring. One is usually interested in the thermodynamic limit when the number of sites $L$ and the number of particles $N$ diverge while the density remains fixed: $L\to\infty$ and $N\to\infty$ with $\rho = N/L$ being fixed. The steady states of the SEP are thus fully characterized by the density: $0<\rho<1$. 

Numerous generalizations of the SEP, partly inspired by applications to protein synthesis \cite{Mac68,LC03,SZL03,Kirone11} and to vehicular traffic \cite{KS99,PS99,AS00,traffic,transport}, have been investigated. These models often involve a facilitation mechanism---the hopping rate depends on more than just the occupancy of the neighboring site~\cite{KS99,PS99,AS00,traffic,transport,J05,S08,BM09,GKR,GR11,RDDD}.  In glassy dynamics, for instance, the particle mobility decreases as the local density increases~\cite{RS03}; the opposite occurs for molecular motors where  a moving particle exerts a hydrodynamic force pushing other particles~\cite{HPLEEE}.  

In extensions of the SEP, the hopping event is determined by the local environment of the hopping particle, e.g., it may depend on occupancies of sites on distance $\leq \ell$ from the hopping particle \cite{K:RP}, where $\ell$ is fixed. Another feature which is always obeyed is that every hopping event involves a single particle. Lattice gases violating this second property have been recently investigated \cite{Zia12,Zia13,Juha13}. In the accelerated exclusion process \cite{Zia12,Zia13}, for instance, the initial hop can trigger at most one more hop: As in the SEP, particles undergo nearest-neighbor hopping, and if a particle hops to a vacant site and joins an island of length $\leq\ell$, the front particle from that island also hops. (An island is a string of occupied sites delimited by vacant sites on both ends.) At most one additional hop is allowed to occur, that is, the second particle cannot trigger another hop. When $\ell=0$, the accelerated exclusion process reduces to the SEP. 

Here we consider the model with no restrictions on island length and avalanche size. Thus if a hopping particle joins an island of arbitrary length, the front particle from this island hops in the same direction, and this second hopping can trigger the third, which can in turn trigger the fourth, etc. ad infinitum. We shall call this model an Exclusion Process with Avalanches (EPA).

One can consider a two-parameter family of models with thresholds both on the island size and on the duration of avalanches: An induced hop occurs only after a particle joins an island of length $\leq \ell$ and the number of induced hops is $\leq a$. In the EPA, whenever a particle joins an island it always triggers the front particle of that island to hop, and an avalanche can be arbitrarily long. Thus $\ell=a=\infty$ for the EPA. Only the extreme versions appear solvable, namely the EPA (as we will demonstrate in this article) and of course the SEP (for which $a=0$ or equivalently $\ell=0$). 

In the next section, we classify the steady states, compute the current for the totally asymmetric EPA, and determine various correlation functions. In Sec.~\ref{asym_EPA:sec} we examine various hydrodynamic solutions, particularly rarefaction and shock waves, arising in the realm of the totally asymmetric EPA. In Sec.~\ref{sym_EPA:sec} we consider the symmetric EPA, compute the density-dependent diffusion coefficient, study the amplitude of self-diffusion, and compare simulation results with theoretical predictions for these transport coefficients. We summarize our results in Sec.~\ref{concl:sec}.

\section{EPA: Steady States}
\label{SS:sec}

\begin{figure}
\centering
\includegraphics[width=10cm]{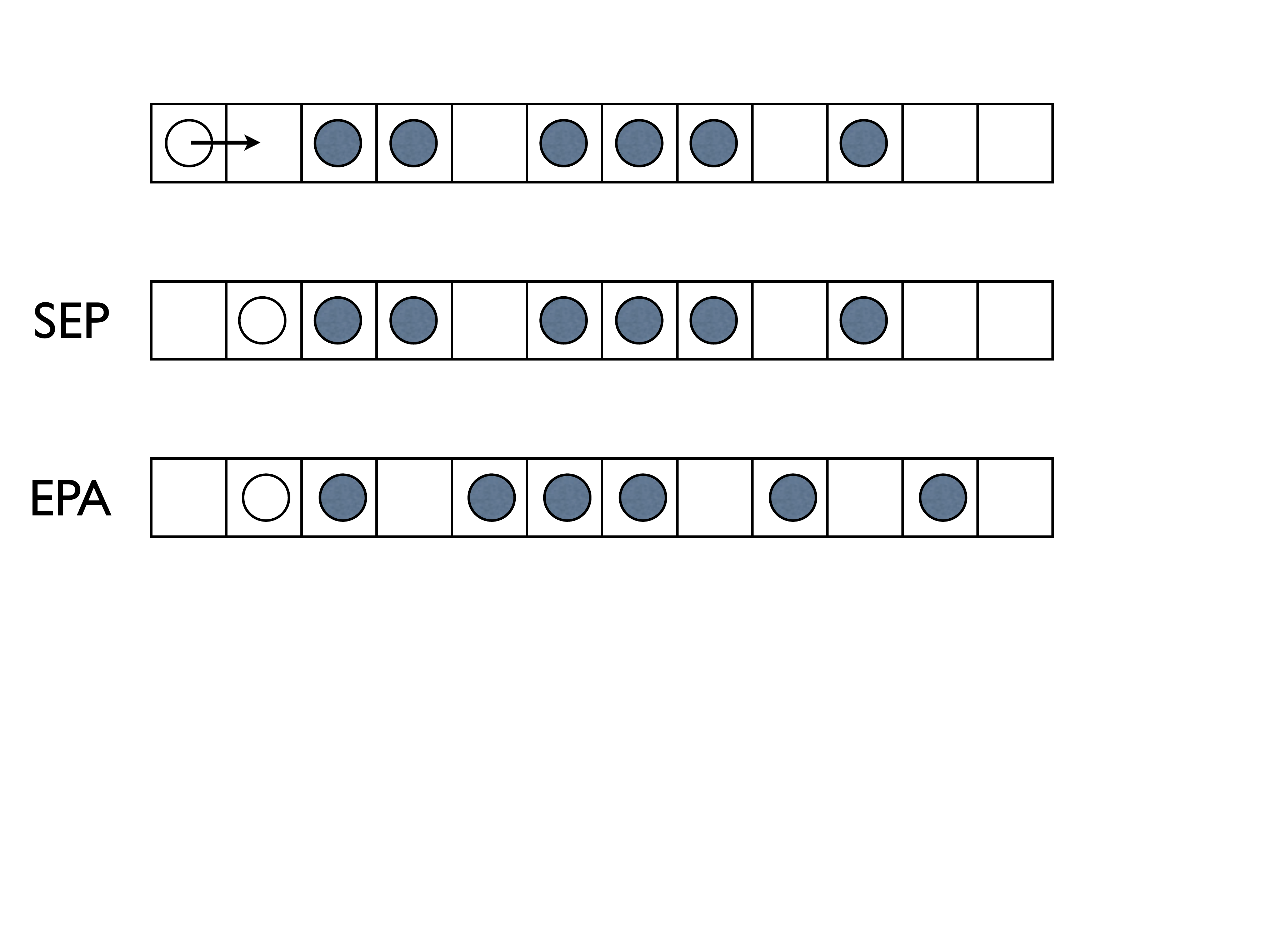}
\vspace{-3cm}
\caption{A hopping event on a ring with 7 particles and 5 empty sites. In this example, a particle (shown as an empty disc) hops to the vacant site on the right. This completes the hopping event in the case of SEP, all other particles (filled discs) remain in their sites. In the case of EPA, the initial hop triggers an avalanche with three induced hops (shown is the final configuration). The initial configuration has 4 islands. After the hopping event there are 3 islands for the SEP and 4 islands for the EPA: In the latter case, the total number of islands cannot decrease.}
\label{Fig:hopping}
\end{figure}

In this section we consider the EPA on the ring. The difference in hopping rules of SEP and EPA is illustrated on Fig.~\ref{Fig:hopping}. The initial configuration is shown on the top. We consider the hopping event which starts when the left-most particle (empty circle) makes the hop to the empty site on the right. This completes the hopping event in the case of SEP (second row). For the EPA (third column), the primary hop triggers the second hop, the second hop triggers the third, which then trigers the fourth, and only then the hopping event is completed as the last hopping particle has not joined an island. 

For the EPA on a finite ring, the memory of the initial condition will be eventually forgotten. We want to understand the nature of the steady states, to determine the current (if the hopping is biased), and to compute density correlations. The high-density regime $\rho>1/2$ is pathological as a never-ending avalanche will eventually occur. (For instance, after a few hopping events, the initial configuration shown on Fig.~\ref{Fig:hopping} enters in an infinite avalanche.) Therefore we tacitly assume that $\rho<1/2$ if not stated otherwise. 

\subsection{Classification of steady states}

The steady states admit a neat classification: They are configurations with the maximal number of islands. Since $\rho<1/2$, the maximal island configurations are such where all islands have length one. Therefore after a transient period,  the EPA reaches a configuration like 
\begin{equation}
\label{example}
\bullet\,\circ\,\circ\,\bullet\,\circ\,\bullet\,\circ\,\bullet\,\circ\,\bullet\,\circ\,\circ\,\circ\,\bullet\,\circ\,\bullet\,\circ\,\bullet\,\circ\,\circ
\end{equation}
and it will then forever wander on the phase space of such maximal-island configurations. In \eqref{example} and other illustrations $\bullet$ denotes a particle and $\circ$ denotes a vacancy. Thus in \eqref{example} we have illustrated a steady state on a ring of length $L=20$ that contains $N=8$ particles and $V=12$ vacancies.

The emergence of the maximal-island configurations is easy to appreciate: After each completed hopping event, the total number of islands increases or remains the same, and in the low-density regime it eventually becomes maximal and then it stays maximal forever. The space of maximal-island configurations is connected: Each maximal-island configuration can evolve into a configuration containing the longest possible string of alternating particles and vacancies complemented by the string of vacancies. 

It turns out that all maximal-island configurations are equally probable. This remarkable property is generally valid for EPA irrespectively is there a bias or not. The totally asymmetric version (say particles hop only to the right) is slightly simpler as it has twice less possible hops than the general EPA, so let's focus on it and derive that maximal-island configurations are equally probable for the totally asymmetric EPA. Apart from expressions for the current [Eqs.~\eqref{current_gen}--\eqref{current:finite} below], the results of this section apply to any EPA. 

Let $P(C)$ be the probability that the system is in maximal-island configuration $C$. In the steady state
\begin{equation}
\label{SS}
P(C)\sum_{C'}R(C\rightarrow C')=\sum_{C''}P(C'')R(C''\rightarrow C)
\end{equation}
where $R(C\rightarrow C')$ is the transition rate from $C$ to $C'$. This rate obeys the zero-one law: $R=1$ if the evolution is allowed and $0$ otherwise. Therefore, we merely need to count the number of ways into and out of a configuration.  Each particle can hop in the maximal-island configuration, so $\sum_{C'}R(C\rightarrow C')=N$. To count the number of maximal-island configurations $C''$ that can change into $C$, we use a simple trick: We reverse the direction of hopping and notice that for each $C''\rightarrow C$ with our original hopping to the right we can find a unique inverse process $C\rightarrow C''$ with hopping to the left. For the latter, the number of ways out is equal to $N$. It must be the same for the former: $\sum_{C''}R(C''\rightarrow C)=N$. If $P(C)$ are equal for all configurations, Eq.~\eqref{SS} is clearly satisfied. 

The probability of a maximum-island configuration is therefore equal to $\mathcal{C}^{-1}$, where $\mathcal{C}$ is the total number of such configurations with $N$ particles and $V$ vacancies that can be arranged on a ring of size $L=N+V$. The total number of maximum-island configurations is
\begin{equation}
\label{conf}
\mathcal{C} = \binom{V}{N} + \binom{V-1}{N-1}
\end{equation}
See \cite{GKR,K:RP} for a computation of a similar quantity. 

\subsection{Current}

In a maximum-island configuration, avalanches are triggered by strings of alternating particles and vacancies. For instance, the snapshot $~\circ\,\circ\,\bullet\,\circ\,\bullet\,\circ\,\bullet\,\circ\,\bullet\,\circ\,\circ~$  represents an alternating string with 4 particles; the illustration \eqref{example} contains the $4-$particle string together with the $3-$particle string and the $1-$particle string. Generally, let $A_k$ be the density of $k-$particle strings. To determine $A_k$ we need to compute the number of configurations where $N-k$ remaining particles are inserted into $V-k-2$ possible positions (denoted by $\downarrow$)
\begin{equation}
\label{Ak:conf} 
\circ\,\circ\,\underbrace{\bullet\,\circ\,\bullet\,\circ\,\bullet\,\circ\,\bullet}_{k~\text{particles}}\,\circ\,\circ\,\overbrace{\downarrow\underbrace{\circ\downarrow\circ\downarrow\circ\downarrow\circ\downarrow\circ\downarrow\circ}_{V-k-3 ~\text{vacancies}}\downarrow}^{N-k~ \text{particles}}
\end{equation}
The example \eqref{Ak:conf} is meant to be general, but what is specifically shown is the alternating string of $k=4$ particles on a ring with total  number of vacancies $V=13$; the total number of particles in  \eqref{Ak:conf} is not specified, although it is bounded $N\leq 11$.

The total number of configurations \eqref{Ak:conf} is $\binom{V-k-2}{N-k}$ and therefore 
\begin{equation}
\label{Ak:finite} 
A_k = \frac{\binom{V-k-2}{N-k}}{\binom{V}{N} + \binom{V-1}{N-1}} 
\end{equation}
This exact result holds independently on the system size. Keeping $k$ finite and going to the thermodynamic limit we find that the density $A_k$ becomes
 \begin{equation}
\label{Ak} 
A_k = \frac{(1-2\rho)^2}{1-\rho}\left(\frac{\rho}{1-\rho}\right)^k
\end{equation}
One computes $\sum_{k\geq 1} kA_k = \rho$ thereby providing a useful consistency check. 
 
Using \eqref{Ak} we can immediately compute the current in the totally asymmetric version of the EPA. 
Each string $A_k$ contributes $1+2+\ldots+k$ once we take into account possible choices of the first hopping particle and resulting avalanches. Therefore
\begin{equation}
\label{current_gen}
J = \sum_{k\geq 1} \frac{k(k+1)}{2}\,A_k
\end{equation}
Using \eqref{Ak} we determine the current in the thermodynamic limit
\begin{equation}
\label{current}
J = 
\begin{cases}
\frac{\rho(1-\rho)}{1-2\rho}    & \rho<\tfrac{1}{2}\\
\infty                                       & \rho\geq \tfrac{1}{2}
\end{cases}
\end{equation}

The current is also infinite on finite rings if $V\leq N$. When $V>N$, the current remains finite. Using \eqref{Ak:finite} and \eqref{current_gen} one can compute the current in this situation
\begin{equation}
\label{current:finite}
J = \frac{VN}{(V-N+1)(V+N)}   
\end{equation}
In particular,
\begin{equation*}
J = N \times
\begin{cases}
\frac{N+1}{4(N+1/2)} &  V = N+1\\
\frac{N+2}{6(N+1)}    &  V = N+2\\
\frac{N+3}{8(N+3/2)} &  V = N+3\\
\frac{N+4}{10(N+2)}  &  V = N+4
\end{cases}
\end{equation*}

\subsection{Correlation functions}

A configuration at time $t$ is fully described by binary variables $n_j(t)$: If the site $j\in\mathbb{Z}$ is empty, $n_j(t)=0$; if it is occupied,  $n_j(t)=1$. The structure of the steady states in the EPA is the same as in a repulsion process for which correlation functions have been recently determined \cite{K:RP}. Using these results we conclude that in the thermodynamic limit the connected pair correlation function is given by
\begin{equation}
\label{nn:simple}
\langle n_i n_j\rangle_c\equiv \langle n_i n_j\rangle -\rho^2 = \rho(1-\rho) \left(-\frac{\rho}{1-\rho}\right)^{|j-i|}
\end{equation}
for all $i$ and $j$. Therefore the connected pair correlation function exhibits a pure exponential decay modulated by an oscillating sign. 

Higher-order correlation functions can be expressed via the pair correlation function, e.g., the three particle correlation function has a neat form
\begin{equation}
\label{3CF}
\langle n_{i_1} n_{i_2} n_{i_3}\rangle = \frac{\langle n_{i_1} n_{i_2}\rangle\,\langle n_{i_2} n_{i_3}\rangle}{\langle  n_{i_2}\rangle}
\end{equation}
reminiscent of the Kirkwood's superposition approximation \cite{Kirkwood35} which is popular in liquid theory \cite{Balescu,RL77}. (Needless to say, for the EAP, and also for the repulsion process studied in  \cite{K:RP},  the above results \eqref{nn:simple}--\eqref{3CF} are exact rather than an uncontrolled approximation.)  Generally the higher-order correlation functions can be written as
\begin{equation}
\label{high_CF}
\left\langle \prod_{a=1}^k n_{i_a}\right\rangle = \frac{1}{\rho^{k-2}}\prod_{a=1}^{k-1}
\left\langle n_{i_a} n_{i_{a+1}}\right\rangle
\end{equation}

\section{Hydrodynamic Solutions}
\label{asym_EPA:sec}

In this section we consider the totally asymmetric EPA. We study evolving solutions, so our setting is the infinite one-dimensional lattice rather than the ring. We are interested in a hydrodynamic description which represents the evolution of the density $\rho(x,t)$ on large spatial and temporal scales. In the hydrodynamic regime, the totally asymmetric EPA is described by the continuity equation with current given by \eqref{current}:
\begin{equation}
\label{AAEP:eq}
\frac{\partial \rho}{\partial t} + \frac{\partial J}{\partial x}  = 0, \quad
J  = \frac{\rho(1-\rho)}{1-2\rho}
\end{equation}

Let us examine solutions which arise when the initial density is a step function
\begin{equation}
\label{step}
\rho = 
\begin{cases}
\rho_- & x<0\\
\rho_+ & x>0
\end{cases}
\end{equation}
There are two types of solutions: Rarefaction waves and shock waves.  

\subsection{Rarefaction and Shock Waves}

We assume that both $\rho_- <\frac{1}{2}$ and $\rho_+ <\frac{1}{2}$, so that the system is in the low-density regime where the current is well-defined. Rarefaction waves arise when $\rho_- <\rho_+ <\frac{1}{2}$. To determine how an initial density step evolves, one can use the method of characteristics \cite{Log94}. The lack of the spatial scale suggests that a simpler approach \cite{book} is to use the scaling ansatz $\rho(x,t)=f(x/t)$.  Plugging this ansatz into Eq.~\eqref{AAEP:eq} and solving the resulting equation we find
\begin{equation}
\label{RW}
f = 
\begin{cases}
\rho_-                                   & z<z_-\\
\tfrac{1}{2}[1-(2z-1)^{-1/2}]  & z_-<z<z_+\\
\rho_+                                  & z>z_+
\end{cases}
\end{equation}
with
\begin{subequations}
\begin{align}
&2z_+ =  1+\frac{1}{(1-2\rho_+)^2}
\label{plus}\\
&2z_- =  1+\frac{1}{(1-2\rho_-)^2}
\label{minus}
\end{align}
\end{subequations}

When $\frac{1}{2}>\rho_- > \rho_+$, the resulting solution is a shock wave. The density profile \eqref{step} translates with velocity
\begin{equation}
\label{v:shock}
v = \frac{J(\rho_+)-J(\rho_-)}{\rho_+ - \rho_- }=\frac{(1-\rho_-)(1-\rho_+) +\rho_-\rho_+}{(1-2\rho_-)(1-2\rho_+)}
\end{equation}
which follows from Eq.~\eqref{AAEP:eq}.

\subsection{Solutions with $\rho_+=1$ or $\rho_-=1$}
\label{exception}

Never-ending avalanches arise when $\frac{1}{2}<\rho<1$, yet the case of $\rho=1$ is non-pathological, it corresponds to the complete stasis. Let us analyze solutions to Eqs.~\eqref{AAEP:eq}--\eqref{step} when the density in one of the half-space is equal to unity. The density profile
\begin{equation}
\label{step_1}
\rho = 
\begin{cases}
\rho_- & x<0\\
1 & x>0
\end{cases}
\end{equation}
with $\rho_- <\frac{1}{2}$ translates with velocity
\begin{equation}
v = -\frac{J(\rho_-)}{1-\rho_-}=-\frac{\rho_-}{1-2\rho_-}
\end{equation}
so we have a shock wave propagating to the left. 

Unexpected results emerge for the complimentary density profile
\begin{equation}
\label{step_2}
\rho = 
\begin{cases}
1 & x<0\\
\rho_+ & x>0
\end{cases}
\end{equation}
When $\rho_+ <\frac{1}{3}$, the solution is a combination of two shock waves
\begin{equation}
\label{SW}
f = 
\begin{cases}
1                                          & z<-1\\
\tfrac{1}{3}                           & -1<z<v\\
\rho_+                                  & z>v
\end{cases}
\end{equation}
where $\rho(x,t)=f(z)$ with $z=x/t$ and $v=\frac{2-\rho_+}{1-2\rho_+}$. To establish this solution we notice that one shock wave moves to the left with unit speed, and if $R$ is the density to the right of this shock wave, equating the mass transfer yields $1-R=J(R)$, from which $R=\frac{1}{3}$ as it is stated in \eqref{SW}. The second shock moves to the right with velocity found from \eqref{v:shock} if we plug in $\rho_-=R=\frac{1}{3}$. The simplest solution of this type describes the expansion into vacuum:
\begin{equation}
\label{SW:vacuum}
f = 
\begin{cases}
1                                          & z<-1\\
\tfrac{1}{3}                           & -1<z<2\\
0                                          & z>2
\end{cases}
\end{equation}

When $\frac{1}{3}<\rho_+ <\frac{1}{2}$, the solution is a combination of a shock wave and a rarefaction wave
\begin{equation}
\label{SW_RW}
f = 
\begin{cases}
1                                          & z<-1\\
\tfrac{1}{3}                           & -1<z<5\\
\tfrac{1}{2}[1-(2z-1)^{-1/2}]  & 5<z<z_+\\
\rho_+                                  & z>z_+
\end{cases}
\end{equation}
The right boundary of the rarefaction wave is determined by Eq.~\eqref{plus}, while the left boundary $z_-=5$ is found after inserting $\rho_-=\frac{1}{3}$ into Eq.~\eqref{minus}.

\section{Symmetric EPA}
\label{sym_EPA:sec}

In the symmetric version, hopping to the left occurs with the same (unit) rate as hopping to the right. Each hopping event can trigger an avalanche propagating in the direction of the initiating hop. Steady states are the same as in the asymmetric version, namely the system wanders on the phase space of the maximum-island configurations and each such configuration occurs with the same probability. The interpretation is different, however: The dynamics is now reversible and the maximum-island configurations are equilibrium configurations since there is no current; mathematically, previous results (e.g., about correlation functions) continue to hold. 

\subsection{Hydrodynamic regime}

To understand the dynamics at a greater depth, one would like to describe the approach to equilibrium.  
Similarly to other lattice gases with reversible dynamics, the hydrodynamic description of the symmetric EPA is provided by the diffusion equation \cite{Spohn91,KL99,book}
\begin{equation}
\label{rho:eq}
\frac{\partial \rho}{\partial t} = \frac{\partial }{\partial x}\!\left[D(\rho)\, \frac{\partial \rho}{\partial x}\right]
\end{equation}

The diffusion coefficient $D(\rho)$ representing the spread of disturbances generically {\em depends} on the density. In rare cases (the SSEP is the most known example) the diffusion coefficient is constant.  Generally, the computation of $D(\rho)$ is very challenging, and a few density-dependent diffusion coefficients have been analytically determined (see e.g. \cite{K:RP}). This is not surprising if we recall that for classical gases transport coefficients cannot be computed even for monoatomic gases with simplest interactions \cite{RL77}. In addition, lattice gases are dense, while classical gases are diluted; for dense classical gases and liquids, the computation of transport coefficients is unimaginable.  Stochastic lattice gases are characterized by a single macroscopic variable, the density, so they are much simpler than classical gases and therefore for some lattice gases the diffusion coefficient is computable.

There is a general scheme based on the Einstein-Green-Kubo formula \cite{RL77} that expresses the diffusion coefficient through the current-current correlation function. Current-current correlations are very difficult to compute for deterministic dense gases. For stochastic lattice gases these correlations are more transparent \cite{Spohn91}, yet successful calculations have been performed in rare cases, mostly for lattice gases satisfying a gradient condition  \cite{Spohn91,MFT_rev}, i.e., when the current can be expressed as a discrete gradient of some function. The EPA does not satisfy the gradient condition, plus the established Einstein-Green-Kubo scheme \cite{Spohn91} assumes single hopping events rather than potential avalanches of simultaneous hops. 

Here we employ an approach which is less general and less justified than the Einstein-Green-Kubo formalism. This approach relies on the knowledge of the steady states and correlations. The calculations are rather involved, but the chief prediction is remarkably simple: 
\begin{equation}
\label{D_EPA}
D = (1-2\rho)^{-3}
\end{equation}
The small-density asymptotic corresponds to the diffusion of a single particle in the empty system and it coincides of course with the diffusion coefficient of the SSEP which is constant: $D_\text{SSEP}=1$. The divergence of the diffusion coefficient in the $\rho\to \tfrac{1}{2}$ limit is expected, although the precise form may be surprising.

To derive \eqref{D_EPA} we assume that the system is already close to equilibrium so that between any two adjacent particles there is at least one vacancy. The density at site $j$ can decrease if the site is occupied and the particle hops to a necessarily empty neighboring site, $j\to j\pm 1$. When a neighboring site is occupied, the particle can hop to the empty site $j$ thereby causing the increase of the density. These direct hops lead to the change of the average density 
\begin{equation}
\label{DN0}
\frac{d \langle n_j\rangle}{dt}\Big|_0 = \langle n_{j-1}\rangle - 2\langle n_j\rangle + \langle n_{j+1}\rangle
\end{equation}
The index on the left-hand side indicates that this change is initiated by direct hopping (no avalanches). 

Similarly we find that the change of the average density due to the first induced hop after the original hop is described by 
\begin{eqnarray}
\label{DN1}
\frac{d \langle n_j\rangle}{dt}\Big|_1 &=& 
\langle n_{j-3} n_{j-1}\rangle - \langle n_{j-2}n_j\rangle\nonumber \\
&-&\langle n_jn_{j+2}\rangle  +\langle n_{j+1}n_{j+3}\rangle 
\end{eqnarray}
Extending this argument we determine the change due to the second induced hop after the original hop 
\begin{eqnarray}
\label{DN2}
\frac{d \langle n_j\rangle}{dt}\Big|_2 &=& 
\langle n_{j-5} n_{j-3} n_{j-1}\rangle - \langle n_{j-4} n_{j-2} n_j\rangle\nonumber \\
&-& \langle n_j n_{j+2} n_{j+4}\rangle + \langle n_{j+1}n_{j+3}n_{j+5}\rangle 
\end{eqnarray}

In the hydrodynamic limit the average density varies on the scales greatly exceeding the lattice spacing. Therefore we write $\langle n_j(t)\rangle =\rho(x,t)$ (the notation $x=j$ emphasizes that we are switching to the continuum description), expand $\langle n_{j\pm 1}\rangle$ in Taylor series 
\begin{equation*}
\langle n_{j\pm 1}\rangle = \rho \pm \rho_x + \tfrac{1}{2}\rho_{xx}+\cdots
\end{equation*}
and recast a difference-differential equation \eqref{DN0} into a classical diffusion equation
\begin{equation}
\label{DE0}
\frac{\partial \rho}{\partial t}\Big|_0 = \frac{\partial^2 \rho}{\partial x^2}
\end{equation}

The right-hand side of Eq.~\eqref{DN1} can be shortly written as $\Psi^{(1)}_{j-2}-\Psi^{(1)}_{j-1}-\Psi^{(1)}_{j+1}+\Psi^{(1)}_{j+2}$, where   $\Psi^{(1)}_k\equiv \langle n_{k-1}n_{k+1}\rangle$. Expanding the right-hand side we transform \eqref{DN1} into 
\begin{equation}
\label{DE1}
\frac{\partial \rho}{\partial t}\Big|_1 = (2^2-1^2)\,\frac{\partial^2 \Psi^{(1)}}{\partial x^2}
\end{equation}

Similarly $\Psi^{(2)}_{j-3}-\Psi^{(2)}_{j-2}-\Psi^{(2)}_{j+2}+\Psi^{(2)}_{j+3}$ with the short-hand notation $\Psi^{(2)}_k\equiv \langle n_{k-2} n_k n_{k+2}\rangle$ is the right-hand side of \eqref{DN2}, so in the continuum limit \eqref{DN2} becomes
\begin{equation}
\label{DE2}
\frac{\partial \rho}{\partial t}\Big|_2 = (3^2-2^2)\,\frac{\partial^2 \Psi^{(2)}}{\partial x^2}
\end{equation}

It is now clear that the general contribution describing the change after the $p^\text{th}$ induced hop is
\begin{equation}
\label{DEp}
\frac{\partial \rho}{\partial t}\Big|_p = [(p+1)^2-p^2]\,\frac{\partial^2 \Psi^{(p)}}{\partial x^2}
\end{equation}
The correlation functions $\Psi^{(p)}$ are direct generalizations of the already defined $\Psi^{(1)}$ and $\Psi^{(2)}$: 
\begin{equation*}
\begin{split}
\Psi^{(3)} &=\langle n_{j-3} n_{j-1} n_{j+1} n_{j+3}\rangle\\
\Psi^{(4)} &=\langle n_{j-4} n_{j-2} n_j n_{j+2} n_{j+4}\rangle\\
\Psi^{(5)} &=\langle n_{j-5} n_{j-3} n_{j-1} n_{j+1} n_{j+3} n_{j+5}\rangle
\end{split}
\end{equation*}
etc. Collecting the contributions from \eqref{DE0} and \eqref{DEp} for all $p\geq 1$ we conclude that the governing hydrodynamic equation reads
\begin{equation}
\label{DER}
\frac{\partial \rho}{\partial t} =  \frac{\partial^2 R}{\partial x^2}\,, \quad 
R = \rho+\sum_{p\geq 1}[(p+1)^2-p^2] \Psi^{(p)}
\end{equation}

We can compute $\Psi^{(p)}$ neglecting the variation of the density.  
Using \eqref{nn:simple} we find 
\begin{equation*}
\Psi^{(1)}  = \langle n_{j-1} n_{j+1}\rangle = \rho^2 + \frac{\rho^3}{1-\rho} = \frac{\rho^2}{1-\rho}
\end{equation*}
which in conjunction with \eqref{high_CF} give us $\Psi^{(p)}$ for all $p\geq 1$:
\begin{equation*}
\Psi^{(p)}  = \left(\frac{\rho}{1-\rho}\right)^{p-1} \Psi^{(1)}
\end{equation*}
Using these results we compute
\begin{equation}
\label{Rr}
R = \rho + \rho\sum_{p\geq 1}(2p+1)\left(\frac{\rho}{1-\rho}\right)^p 
= \frac{\rho-\rho^2}{(1-2\rho)^2}
\end{equation}

Equation \eqref{DER} can be re-written as the diffusion equation \eqref{rho:eq} with $D(\rho)=\frac{dR}{d\rho}$. Combining this relation with \eqref{Rr} we arrive at the announced diffusion coefficient \eqref{D_EPA}. 

We emphasize that whenever the predictions of the above perturbative approach were compared with rigorous derivations available for a few lattice gases satisfying the gradient condition, e.g., for repulsion processes \cite{K:RP}, the results fully agreed. For lattice gases of non-gradient type the perturbative approach also apparently gives exact results, although the supporting evidence is mostly numerical  (see e.g. \cite{CKM}). 

\subsection{Self-diffusion}

The phenomenon of self-diffusion refers to the evolution of a tagged particle. Self-diffusion is interesting when a lattice gas is at equilibrium (as we shall assume in this subsection), and it can be studied for an arbitrary lattice gas in arbitrary spatial dimension $d$. The average position of the tagged particle remains constant
\begin{equation}
\label{av_tag}
\langle {\bf X}(t)\rangle = {\bf 0}
\end{equation}
The most interesting information is provided by the mean-square displacement. One anticipates that it exhibits a diffusive growth
\begin{equation}
\label{var_tag}
\langle {\bf X}^2(t)\rangle = 2d D_T(\rho)\,t
\end{equation}
The coefficient of self-diffusion $D_T(\rho)$ is unknown even for simplest lattice gases, e.g. for the SSEP in dimensions $d\geq 2$. (Generally, the diffusion of the tagged particle in higher dimensions is described by the self-diffusion matrix, so one should replace \eqref{var_tag} by an obvious matrix generalization.)

In one dimension, the coefficient of self-diffusion may vanish. It happens for all exclusion processes with symmetric hopping when no more than one particle per site and only nearest-neighbor jumps are allowed. For such exclusion processes the mean-square displacement exhibits a sub-diffusive growth. This was originally discovered for the SSEP where the mean-square displacement grows as \cite{Harris_65,Levitt_73,PMR_77,AP_78,A_83,van_83}
\begin{equation}
\label{var_tag_1d}
\langle X^2(t)\rangle = \mathcal{D}(\rho)\,\sqrt{t}\,, \quad 
\mathcal{D}_\text{SSEP}(\rho)=\frac{2}{\sqrt{\pi}}\,\frac{1-\rho}{\rho}
\end{equation}
Note that as in the general case of normal self-diffusion, $P(X,t)=\text{Prob}(X(t)=X)$ is a Gaussian distribution characterized by the average \eqref{av_tag} and the variance \eqref{var_tag_1d}. The anomalously slow growth of the variance, $\sqrt{t}$ instead of the usual linear growth, is caused by the fact that the original order of all the particles is forever preserved in one dimension. The sub-diffusive growth law \eqref{var_tag_1d} is not merely an outcome of a toy model, it has been observed in a number of experimental realizations such as diffusion of large molecules in zeolites,  transport in super-ionic conductors, etc., see \cite{karger,chou,wei} and references therein. 

The $\sqrt{t}$ growth of the variance should be valid for other one-dimensional exclusion processes with symmetric nearest-neighbor hopping. The amplitude $\mathcal{D}(\rho)$ generally depends on the details of the process. The derivation of \eqref{var_tag_1d}, see e.g. \cite{AP_78,van_83,PKT_14}, suggests that $\mathcal{D}(\rho)$ can be expressed through the diffusion coefficient $D(\rho)$ and the static compressibility $\chi(\rho)$ via
\begin{equation}
\label{var_tag_conj}
\mathcal{D}(\rho)= \frac{2}{\sqrt{\pi}}\,\frac{\chi(\rho)}{\rho^2}\,\sqrt{D(\rho)}
\end{equation}
The static compressibility (also known as the structure factor) can be expressed through the connected pair correlation function \cite{Spohn91}
\begin{equation*}
\chi = \sum_{j=-\infty}^\infty \langle n_0n_j\rangle_c
\end{equation*}
Using \eqref{nn:simple} we compute the static compressibility
\begin{equation}
\label{chi:EPA}
\chi(\rho) = \rho(1-\rho)(1-2\rho)
\end{equation}

The range of applicability of \eqref{var_tag_conj} is not fully understood. It is proved to be correct for gradient lattice gases, but the EPA fails this test, plus all rigorous work disregards avalanches (only one jump can occur in an infinitesimal time interval). On the other hand, the validity of Eq.~\eqref{var_tag_conj} has been recently justified \cite{PKT_14} for a general class of lattice gases with exclusion constraint; this has been done in the framework of the macroscopic fluctuation theory (see \cite{MFT_rev} for a review).  
With all these caveats, we now substitute \eqref{D_EPA} and \eqref{chi:EPA} into \eqref{var_tag_conj} and arrive at
\begin{equation}
\label{selfD_EPA}
\mathcal{D}(\rho)=\frac{2}{\sqrt{\pi}}\,\frac{1-\rho}{\rho\sqrt{1-2\rho}}
\end{equation} 

The small density behavior of $\mathcal{D}(\rho)$ matches the behavior in the case of SSEP. The divergence of $\mathcal{D}(\rho)$ when $\rho\rightarrow 1/2$ is also natural. Note that the behavior of $\mathcal{D}(\rho)$ in this limit is less singular than the behaviors of $J(\rho)$ and $D(\rho)$, see \eqref{current} and \eqref{D_EPA}. The amplitude of self-diffusion \eqref{selfD_EPA} is minimal, $\mathcal{D}_\text{min}=3.75771778\ldots$, at $\rho=(3-\sqrt{5})/2= 0.381966012\ldots$. 

\subsection{Probing the diffusion coefficient and the amplitude of self-diffusion}

One can determine the diffusion coefficient numerically by measuring the average flux $\langle F\rangle$, namely the average number of particles passing  through the system of size $L$ during time $t$: One sets the density on the left boundary to $\rho$ and the density on the right boundary to $\rho - \delta \rho$, assumes $\delta\rho\ll \rho$ and $L\gg 1$, and employs relation
\begin{equation*}
\lim_{t\to\infty} \frac{1}{t}\,\langle F\rangle = D(\rho)\,\frac{\delta\rho}{L}
\end{equation*}
to probe the diffusion coefficient. This direct method requires a long running time since the average flux is proportional to $\delta \rho$, while fluctuations of the flux remain finite even for $\delta \rho=0$.

We employ a less direct way of probing the diffusion coefficient which has an advantage of being resilient towards fluctuations. The idea is to consider stationary density profiles with sufficiently different boundary densities. We can compare the density profile observed numerically with the one found theoretically using the predicted expression \eqref{D_EPA} for the diffusion coefficient. 

\begin{figure}[H]
\centering
\includegraphics[width=8.64cm]{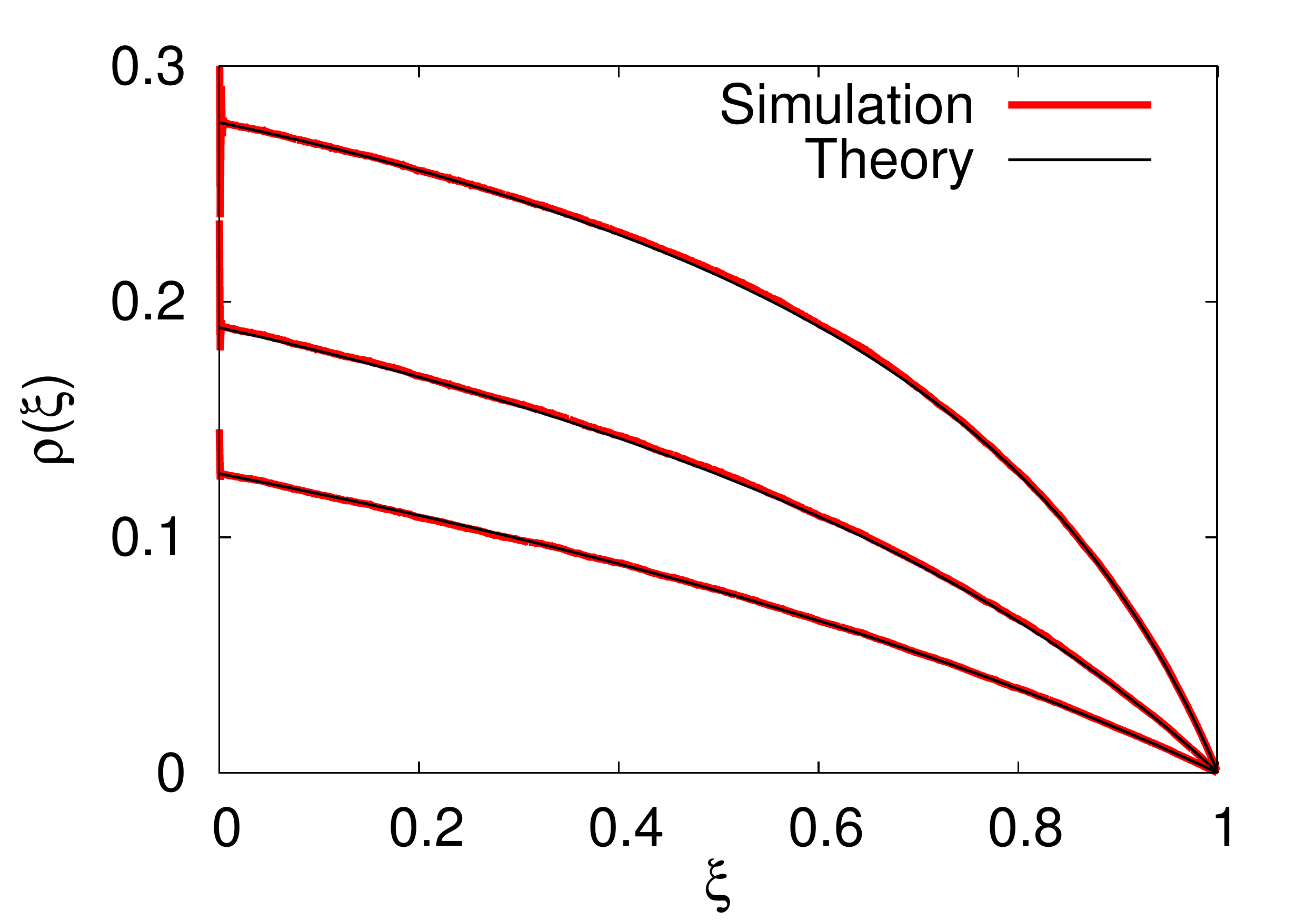}
\vspace{-0.5cm}
\caption{Density profiles: $\rho$ versus the scaled spatial coordinate $\xi=x/L$. Shown are simulation
results for the system with $L=10^3$ and $T=10^8$, for three different densities on the left. Also shown are theoretical predictions, Eq.~\eqref{density_profile}. }
\label{Fig:diffusion-profile}
\end{figure}

For concreteness, we choose the boundary conditions
\begin{equation}
\label{BC}
\rho(0) = \rho_0,  \quad \rho(L) = 0
\end{equation}
Solving \eqref{rho:eq}--\eqref{D_EPA} subject to the boundary conditions \eqref{BC} yields the density profile,
\begin{equation}
\label{density_profile}
2\rho = 1 -  \frac{1-2\rho_0}{\sqrt{1-4\rho_0(1-\rho_0) \xi}}\,, \quad \xi = \frac{x}{L}
\end{equation}
which is valid for any $\rho_0 < 1/2$. Let us compare this to the density profile given by direct simulations of the EPA on the interval $(0,L)$. To achieve the boundary conditions \eqref{BC}, a particle at the site $L-1$ hops to the right with the same rate 1 as in the bulk, but there are no hopping from site $L$ to site $L-1$. And we add particles to site 1 at rate $\lambda$. Note that the introduction of a particle at site 1 will induce an avalanche if there is a particle at the second site. This causes the first few sites to behave differently than the bulk for which we derive the hydrodynamic equations: The first site has a higher density than the macroscopic prediction; the second site has a lower density due to avalanche induced by teh injection of particles at the first site; the third site has a higher density; and so on. These oscillations die out quickly and we fit $\rho_0$ to extrapolated bulk density at $x=0$. In the bulk, there is an excellent agreement between simulations and theory (see Fig.~\ref{Fig:diffusion-profile}).

\begin{figure}[H]
\centering
\includegraphics[width=8.64cm]{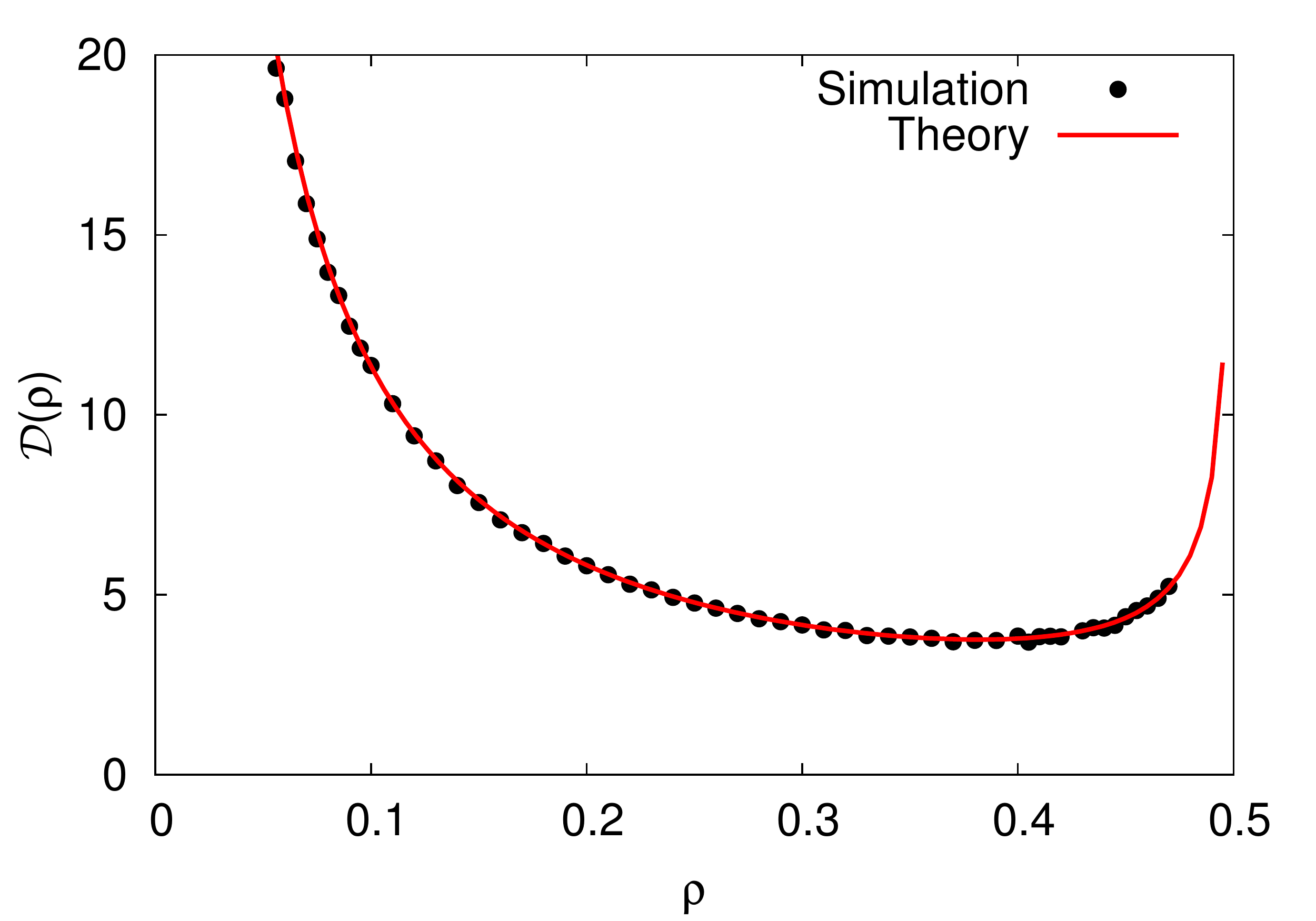}
\vspace{-0.5cm}
\caption{The amplitude of self-diffusion $\mathcal{D}(\rho)$ in one dimension as a function of density $\rho$.  Shown are simulation results on the ring of length $L=10^4$ for $T=10^4$; the averaging was taken over $10^3$ configurations for each $\rho$. Also shown for reference is the theoretical prediction, Eq.~\eqref{selfD_EPA}.}
\label{Fig:selfdiffusion}
\end{figure}

We numerically study the self-diffusion process on a ring. Simulations on the ring provide a faithful description of the infinite lattice as long as the observation time is sufficiently short, $T\ll L^2$. We average the squared displacements over many configurations. One can also sample multiple particle-displacements in the same configuration, while making sure that particles are chosen far enough that the correlations are minimal. [In our simulations we chose particles such that the correlations are $<0.005$. The spacings between samples can be estimated by using \eqref{nn:simple}.] Figure \ref{Fig:selfdiffusion} shows a good agreement between simulations and theory. 

\section{Summary}
\label{concl:sec}

We introduced and investigated exclusion processes with avalanches. In these processes in addition to the hopping to neighboring empty sites characterizing simple exclusion processes, simultaneous hops, the avalanches, can occur. An avalanche is generated when a particle hops to an empty site which is adjacent to an island---in this case, the frontmost particle from this island hops and if this particle joins another island this triggers another hop, etc. There are no restrictions on the length of islands and the duration of the avalanche. Exclusion processes with avalanches are well-defined in the $\rho<\frac{1}{2}$ region. Avalanches lead to an accelerated phenomenon in the sense that both the current and the diffusion coefficient are convex function of the density increasing in the $0<\rho<\frac{1}{2}$ region and diverging in the $\rho\to \frac{1}{2}$ limit. The amplitude of self-diffusion is also a convex function of density diverging in the $\rho\to 0$ and $\rho\to \frac{1}{2}$ limits. 

We showed that for one-dimensional exclusion processes with avalanches the steady states are configurations with maximal number of islands and all these configurations are equiprobable. This understanding allowed us to employ a combinatorial approach to compute the current (for the asymmetric version) and the correlation functions. In the asymmetric version, the continuity equation governs hydrodynamic behaviors, and we determined some key hydrodynamic solutions, particularly rarefaction and shock waves. For the symmetric version, the hydrodynamic behavior is governed by diffusion equation. We computed the density-dependent diffusion coefficient. We also studied a phenomenon of self-diffusion. As in other exclusion processes in one dimension with nearest-neighbor hopping, the mean square displacement of a tracer particle grows as $\mathcal{D}(\rho)\sqrt{t}$ rather than linearly in time as in normal diffusion. We determined the amplitude of self-diffusion $\mathcal{D}(\rho)$. The predicted values of the diffusion coefficient and $\mathcal{D}(\rho)$ are in good agreement with simulation results.

Thus one-dimensional simple exclusion processes with avalanches of unlimited durations are tractable, e.g., transport coefficients exhibit non-trivial density dependence yet they are computable. Exclusion processes with avalanches of limited duration constitute an obvious challenge, some of these processes have been investigated in Refs.~\cite{Zia12,Zia13,Juha13}, but they haven't been solved so far. 

The one-dimensional exclusion process with avalanches has intriguing similarities with lattice gas models without avalanches.  One interesting example is an exclusion process in which particles undergo long `frog-leaping' jumps. This lattice gas has been studied in the context of the self-organized criticality \cite{SOC:sing}. It would be interesting to explore these similarities further, as well as the potential connections with other lattice gases and with zero range processes. Another promising direction is to devise higher-dimensional exclusion processes with avalanches. Naive generalizations appear ill-defined for arbitrarily low densities due to the emergence of never-ending avalanches. 

\medskip
\textbf{Acknowledgments.}\quad 
This work was partially suported by NSF Grant No. DMR-1205797 and BSF Grant No. 2012145.  We are grateful to D. Gabrielli, J. Krug and C. Landim for discussions and suggestions.


\begin{thebibliography}{99}

\bibitem{K66}
     K. Kawasaki, Phys. Rev. {\bf 145}, 224 (1966).

\bibitem{Spohn91}   
     H. Spohn, {\it Large Scale Dynamics of Interacting Particles} 
     (New York: Springer-Verlag, 1991).    

\bibitem{SZ95} 
     B. Schmittmann and R. K. P. Zia, Statistical Mechanics of
     Driven Diffusive Systems, in {\it Phase Transitions and Critical
     Phenomena}, Vol.\ 17, eds.\ C. Domb and J. L. Lebowitz (Academic Press, London).

\bibitem{D98} B. Derrida, Phys.\ Repts.\ {\bf 301}, 65 (1998).

\bibitem{KL99} 
     C. Kipnis and C. Landim, {\it Scaling Limits of Interacting Particle Systems}
     (Springer, New York,  1999).

\bibitem{S00} 
     G. M. Sch\"utz, Exactly Solvable Models for Many-Body Systems
     Far From Equilibrium, in {\it Phase Transitions and Critical Phenomena},
     Vol.\ 19, eds.\ C. Domb and J. L. Lebowitz (Academic Press, London, 2000).

\bibitem{BE07} 
     R. A. Blythe and M. R. Evans, J. Phys.\ A {\bf 40}, R333 (2007).

\bibitem{D07}
     B. Derrida, J. Stat. Mech. P07023 (2007).

\bibitem{book} 
     P. L. Krapivsky, S. Redner, and E. Ben-Naim, {\it A Kinetic
     View of Statistical Physics} (Cambridge University Press, Cambridge,  2010).

\bibitem{Mac68} 
    C. MacDonald, J. Gibbs, and A. Pipkin, Biopolymers {\bf 6}, 1 (1968);
    C. MacDonald and J. Gibbs, Biopolymers {\bf 7,} 707 (1969).

\bibitem{LC03} G. Lakatos and T. Chou, J. Phys.\ A {\bf 36}, 2027 (2003).

\bibitem{SZL03} 
     L. B. Shaw, R. K. P. Zia, and K. H. Lee, Phys.\ Rev.\ E {\bf 68}, 021910 (2003).

\bibitem{Kirone11}
     T. Chou, K. Mallick, and R. K. P. Zia, Rep. Prog. Phys. {\bf 74}, 116601 (2011).
     
\bibitem{KS99} 
     K. Klauck and A. Schadschneider, Physica A {\bf 271}, 102 (1999).

\bibitem{PS99} 
     V. Popkov and G.M. Sch\"utz,  Europhys.\ Lett.\ {\bf 48}, 257 (1999). 

\bibitem{AS00} T. Antal and G. M. Sch\"utz, Phys.\ Rev.\ E {\bf 62}, 83 (2000).

\bibitem{traffic}
     D. Chowdhury, L. Santen, and A. Schadschneider,
     Phys. Reports {\bf 329}, 199 (2000).

\bibitem{transport}
     A. Schadschneider, D. Chowdhury, and K. Nishinari,
     {\it Stochastic Transport in Complex Systems} (Elsevier Science, 2010).

\bibitem{J05} K. Jain, Phys.\ Rev.\ E {\bf 72}, 017105 (2005).

\bibitem{S08} M.~Sellitto, Phys.\ Rev.\ Lett.\ {\bf 101}, 048301 (2008).

\bibitem{BM09} U. Basu and P. K. Mohanty,  Phys.\ Rev.\ E {\bf 79}, 041143 (2009).

\bibitem{GKR} 
     A. Gabel, P. L. Krapivsky, and S. Redner, Phys.\ Rev.\ Lett. {\bf 105}, 210603 (2010). 

\bibitem{GR11} 
     A. Gabel and S. Redner, J. Stat. Mech. P06008 (2011). 

\bibitem{RDDD} 
    R. Dandekar and D. Dhar, EPL {\bf 104}, 26003 (2013). 

\bibitem{RS03} 
     A review of this topic is given in F. Ritort and P. Sollich, Adv.\ Phys.\ {\bf 52}, 219 (2003).

\bibitem{HPLEEE} 
    D. Houtman, I. Pagonabarraga, C. P.  Lowe, A. Esseling-Ozdoba, A. M. C. Emons, and E. Eiser, 
    Europhys.\ Lett.\ {\bf 78}, 18001 (2007).

\bibitem{K:RP} 
     P. L. Krapivsky, J. Stat. Mech. P06012 (2013). 
     
\bibitem{Zia12}
     J. Dong, S. Klumpp, and R. K. P. Zia,  Phys. Rev. Lett. {\bf 109}, 130602 (2012).
     
\bibitem{Zia13}
     J. Dong, S. Klumpp, and R. K. P. Zia,  Phys. Rev. E {\bf 87}, 022146 (2013).

\bibitem{Juha13}
     J. Merikoski, Phys. Rev. E {\bf 88}, 062137 (2013).

\bibitem{Kirkwood35}        
     J. G. Kirkwood, J. Chem. Phys. {\bf 3}, 300 (1935).
       
\bibitem{Balescu}       
     R. Balescu, {\it Equilibrium and Nonequilibrium Statistical Mechanics} (Wiley, New York, 1975).

\bibitem{RL77} 
     P. Resibois and M. De Leener, {\it Classical Kinetic Theory of Fluids} (Wiley, New York, 1977).

\bibitem{Log94} 
     D. J. Logan, {\it An Introduction to Nonlinear Partial
     Differential Equations} (Wiley, New York, 1994).

\bibitem{MFT_rev}  
     L. Bertini, A. De Sole, D. Gabrielli, G. Jona-Lasinio, and C. Landim, arXiv:1404.6466.

\bibitem{CKM}
    C. Arita, P. L. Krapivsky, and K. Mallick, arXiv:1407.3228. 

\bibitem{Harris_65} 
     T. E. Harris, J. Appl. Prob. {\bf 2}, 323 (1965).

\bibitem{Levitt_73} 
    D. G. Levitt, Phys. Rev. A {\bf 8}, 3050 (1973). 

\bibitem{PMR_77}
    P. M. Richards, Phys. Rev. B {\bf 16}, 1393 (1977). 

\bibitem{AP_78} 
    S. Alexander and P. Pincus, Phys. Rev. B {\bf 18}, 2011 (1978).

\bibitem{A_83}
    R. Arratia, Ann. Probab. {\bf 11}, 362 (1983).  

\bibitem{van_83} 
    H. van Beijeren, K. W. Kehr, and R. Kutner, Phys. Rev. B {\bf 28}, 5711 (1983).
 
\bibitem{karger}
     J. K\"{a}rger and D. Ruthven, {\it Diffusion in Zeolites and Other Microporous Solids}
     (Wiley, New York, 1992).
     
\bibitem{chou}
     T. Chou and D. Lohse, Phys. Rev. Lett. {\bf 82}, 3552 (1999).         

\bibitem{wei}
     Q.-H. Wei, C. Bechinger, and P. Leiderer, Science {\bf 287}, 625 (2000); 
     C. Lutz, M. Kollmann, and C. Bechinger, Phys. Rev. Lett. {\bf 93}, 026001 (2004).

\bibitem{PKT_14} 
     P. L. Krapivsky, K. Mallick, and T. Sadhu, arXiv:1405.1014.

\bibitem{SOC:sing}
   J. M. Carlson, J. T. Chayes, E. R. Grannan, and G. H. Swindle, Phys. Rev. Lett. {\bf 65}, 2547 (1990). 


\end{thebibliography}
\end{document}